\documentclass{revtex4}
\usepackage{amsmath}

\input{tcilatex}
\begin{document}

\preprint{}
\title{Quantum singularities in ($2+1)$ dimensional matter coupled black
hole spacetimes }
\author{O. Unver}
\author{O. Gurtug}
\email{ozlem.unver@emu.edu.tr}
\email{ozay.gurtug@emu.edu.tr}
\affiliation{Department of Physics, Eastern Mediterranean University, G. Magusa, North
Cyprus, Mersin 10 - Turkey.}

\begin{abstract}
Quantum singularities considered in the $3D$ BTZ spacetime by Pitelli and
Letelier (Phys. Rev. D77: 124030, 2008) is extended to charged BTZ and 3D
Einstein-Maxwell-dilaton gravity spacetimes. The occurence of naked
singularities in the Einstein-Maxwell extension of the BTZ spacetime both in
linear and non-linear electrodynamics as well as in the
Einstein-Maxwell-dilaton gravity spacetimes are analysed with the quantum
test fields obeying the Klein-Gordon and Dirac equations. We show that with
the inclusion of the matter fields; the conical geometry near $r=0$ is
removed and restricted classes of solutions are admitted for the
Klein-Gordon and Dirac equations. Hence, the classical central singularity
at $r=0$ turns out to be quantum mechanically singular for quantum particles
obeying Klein-Gordon equation but nonsingular for fermions obeying Dirac
equation. Explicit calculations reveal that the occurrence of the timelike
naked singularities in the considered spacetimes do not violate the cosmic
censorship hypothesis as far as the Dirac fields are concerned. The role of
horizons that clothes the singularity in the black hole cases is replaced by
repulsive potential barrier against the propagation of Dirac fields.
\end{abstract}

\maketitle

\section{INTRODUCTION}

In recent years, $\left( 2+1\right) $ dimensional,
Banados-Teitelboim-Zanelli (BTZ) \cite{BZ} black hole has attracted much
attention. One of the basic reasons for this attraction is that the BTZ
black hole has a relatively simple tractable mathematical structure so that
it provides a better understanding of investigating the general aspects of
black hole physics since the BTZ black hole carries all the characteristic
features such as the event horizon and Hawking radiation, it can be treated
as a real black hole. Another motivation to study BTZ black hole is the
AdS/CFT correspondence which relates thermal properties of black holes in
the AdS space to a dual CFT. In view of these points, the unresolved black
hole properties belonging to $\left( 3+1\right) $ or higher dimensional
black holes at the quantum level, make the BTZ black hole an excellent
background for exploring the black hole physics.

Another interesting subject is the study of naked singularities that can be
considered as a threat to the cosmic censorship hypothesis. Compared to the
black holes, the naked singularities are less understood. Today, there is no
common consensus either on the structure or the existence of the naked
singularities.

Recently, Pitelli and Letelier (PL) \cite{PL} have analysed the occurrence
of naked singularities for the BTZ spacetime from quantum mechanical point
of view. In their analysis, the criteria proposed by Horowitz and Marolf
(HM) \cite{HM} is used. The classical naked singularity is studied with the
quantum test particles that obey Klein-Gordon and Dirac equations. They
confirmed that the naked singularity is "healed" when tested by massless
scalar particles or fermions without introducing extra boundary conditions.
However, for massive scalar particles additional information is needed.
Despite the recent developments on the concept of quantum singularities \cite%
{HK}, our understanding of naked singularities as far as quantum gravity
concerned is still far from being complete.

The purpose of this paper is to analyse the naked singularities within the
context of the quantum mechanics that form in the matter coupled $2+1$
dimensional black hole spacetimes. Our motivation here is to investigate the
effect of the matter fields on the quantum singularity structure of the BTZ
spacetime because the surface at $r=0$ for the BTZ black hole is not a
curvature singularity, but is a singularity in the causal structure. This
situation changes when a matter field is coupled. This is precisely the case
that we shall elaborate on in this article. For this purpose we consider the
charged BTZ spacetime both in linear and non-linear electrodynamics. This is
analogous to a kind of Einstein-Maxwell extension of the work presented in 
\cite{PL}. Furthermore, we extend the analysis to cover $2+1$ dimensional
Einstein-Maxwell-dilaton coupled black hole spacetime. The presence of
charge both in linear and non-linear case and also the dilaton field modify
the resulting spacetime geometry significantly. Near the origin, the
spacetime is not conic and true curvature singularity develops at $r=0$.
Consequently, the spacetime geometry that we have investigated in this study
differs when compared with the case considered in \cite{PL}.

The plan of the paper is as follows. In section II, we first review the
definition of quantum singularities for general static spacetimes. In
section III, we consider the charged BTZ black hole in non-linear
electrodynamics. Klein-Gordon and Dirac fields are used to test the quantum
singularity. We also discuss the Sobolev norm which is used first time in
this context by Ishibashi and Hosoya \cite{IH}. In sections IV and V, we
consider charged BTZ in linear electrodynamics and dilaton coupled $3D$
black hole spacetime in the Einstein-Maxwell and Einstein - Maxwell -
dilaton theory, respectively. Dirac and scalar fields are used to judge the
quantum singularity. The paper ends with a conclusion in section VI.

\section{A Brief Review of Quantum Singularities}

In classical general relativity, the spacetime is said to be singular if the
evolution of timelike or null geodesics is not defined after a proper time.
Horowitz and Marolf, \ based on the pioneering work of Wald \cite{WA}, have
proposed the criteria to test the classical singularities with quantum test
particles that obey the Klein-Gordon equation for static spacetime having
timelike singularities. According to this criteria, the singular character
of the spacetime is defined as the ambiguity in the evolution of the wave
functions. That is to say, the singular character is determined in terms of
the ambiguity when attempting to find self-adjoint extension of the operator
to the entire space. If the extension is unique, it is said that the space
is quantum mechanically regular. The brief review is as follows:

Consider a static spacetime $\left( M,g_{\mu\nu}\right) $\ with a timelike
Killing vector field $\xi^{\mu}$. Let $t$ denote the Killing parameter and $%
\Sigma$\ denote a static slice.The Klein-Gordon equation on this space is

\begin{equation}
\left( \nabla^{\mu}\nabla_{\mu}-M^{2}\right) \psi=0.
\end{equation}
This equation can be written in the form of

\begin{equation}
\frac{\partial^{2}\psi}{\partial t^{2}}=\sqrt{f}D^{i}\left( \sqrt{f}%
D_{i}\psi\right) -fM^{2}\psi=-A\psi,
\end{equation}
in which $f=-\xi^{\mu}\xi_{\mu}$ and $D_{i}$ is the spatial covariant
derivative on $\Sigma$. The Hilbert space $\left( L^{2}\left( \Sigma\right)
\right) $\ is the space of square integrable functions on $\Sigma$. The
domain of the operator $A,$ $D(A)$ is taken in such a way that it does not
enclose the spacetime singularities. An appropriate set is $C_{0}^{\infty
}\left( \Sigma\right) $, the set of smooth functions with compact support on 
$\Sigma$. Operator $A$ is real, positive and symmetric therefore its
self-adjoint extensions always exist. If \ it has a unique extension $A_{E},$
then $A$ is called essentially self-adjoint \cite{RS}. Accordingly, the
Klein-Gordon equation for a free particle satisfies

\begin{equation}
i\frac{d\psi}{dt}=\sqrt{A_{E}}\psi,
\end{equation}
with the solution

\begin{equation}
\psi\left( t\right) =\exp\left[ -it\sqrt{A_{E}}\right] \psi\left( 0\right) .
\end{equation}
If $A$ is not essentially self-adjoint, the future time evolution of the
wave function ( Eq. (4)) is ambiguous. Then, Horowitz and Marolf define the
spacetime quantum mechanically singular. However, if there is only one
self-adjoint extension, the operator $A$ is said to be\ essentially
self-adjoint and the quantum evolution described by Eq.(4) is uniquely
determined by the initial conditions. According to the Horowitz and Marolf
criterion, this spacetime is said to be quantum mechanically nonsingular. In
order to determine the number of self-adjoint extensions, the concept of
deficiency indices is used. The deficiency subspaces $N_{\pm}$ are defined
by ( see Ref. \cite{IH} for a detailed mathematical background),

\begin{align}
N_{+} & =\{\psi\in D(A^{\ast}),\text{ \ \ \ \ \ \ }A^{\ast}\psi=Z_{+}\psi,%
\text{ \ \ \ \ \ }ImZ_{+}>0\}\text{ \ \ \ \ \ with dimension }n_{+} \\
N_{-} & =\{\psi\in D(A^{\ast}),\text{ \ \ \ \ \ \ }A^{\ast}\psi=Z_{-}\psi,%
\text{ \ \ \ \ \ }ImZ_{-}>0\}\text{ \ \ \ \ \ with dimension }n_{-}  \notag
\end{align}
The dimensions $\left( \text{ }n_{+},n_{-}\right) $ are the deficiency
indices of the operator $A$. The indices $n_{+}(n_{-})$ are completely
independent of the choice of $Z_{+}(Z_{-})$ depending only on whether $Z$
lies in the upper (lower) half complex plane. Generally one takes $%
Z_{+}=i\lambda$ and $Z_{-}=-i\lambda$ , where $\lambda$ is an arbitrary
positive constant necessary for dimensional reasons. The determination of
deficiency indices then reduces to counting the number of solutions of $%
A^{\ast}\psi=Z\psi$ ; (for $\lambda=1$),

\begin{equation}
A^{\ast}\psi\pm i\psi=0
\end{equation}
that belong to the Hilbert space $\mathcal{H}$. If there is no square
integrable solutions ( i.e. $n_{+}=n_{-}=0)$, the operator A possesses a
unique self-adjoint extension and it is essentially self-adjoint.
Consequently, a sufficient condition for the operator $A$ to be essentially
self-adjoint is to investigate the solutions satisfying Eq. (6) that do not
belong to the Hilbert space.

\section{$\left( 2+1\right) $ - Dimensional BTZ Spacetime Coupled with
Non-linear Electrodynamics}

\subsection{Solutions and Spacetime Structure:}

The action describing $(2+1)$ - dimensional Einstein theory coupled with
non-linear electrodynamics is given by \cite{MC},

\begin{equation}
S=\int\sqrt{g}\left( \frac{1}{16\pi}\left( R-2\Lambda\right) +L(F)\right)
d^{3}x.
\end{equation}
The field equations via variational principle read as,

\begin{equation}
G_{ab}+\Lambda g_{ab}=8\pi T_{ab},
\end{equation}

\begin{equation}
T_{ab}=g_{ab}L(F)-F_{ac}F_{b}^{\text{ \ }c}L_{,F},
\end{equation}

\begin{equation}
\nabla_{a}\left( F^{ab}L_{,F}\right) =0
\end{equation}
in which $L_{,F}$ stands for the derivative of $L(F)$ with respect to $F=%
\frac{1}{4}F_{ab}F^{ab}$. The non-linear field is chosen so that the energy
momentum tensor (9) has a vanishing trace. The trace of the tensor gives,

\begin{equation}
T=T_{ab}g^{ab}=3L(F)-4FL_{,F}.
\end{equation}
Hence, to have a vanishing trace, the electromagnetic Lagrangian is obtained
as

\begin{equation}
L=c\mid F\mid^{3/4},
\end{equation}
where $c$ is an integration constant. With reference to the paper \cite{MC},
the complete solution to the above action is given by the metric,

\begin{equation}
ds^{2}=-f(r)dt^{2}+f(r)^{-1}dr^{2}+r^{2}d\theta^{2},
\end{equation}
where the metric function $f(r)$ is given by,

\begin{equation}
f(r)=-m+\frac{r^{2}}{l^{2}}+\frac{4q^{2}}{3r}.
\end{equation}
Here $m>0$ is the mass, $l^{2}=-\Lambda^{-1}$ the case $\Lambda>0$ $%
(\Lambda<0)$, that corresponds with an asymptotically de-Sitter (anti
de-Sitter) spacetime, and $q$ is the electric charge. This metric represents
the BTZ spacetime in non-linear electrodynamics. If $\Lambda=0$, we have an
asymptotically flat solution coupled with Coulomb-like field. The
Kretschmann scalar which indicates the occurrence of curvature singularity
is given by,

\bigskip

\begin{equation}
\mathcal{K}=\frac{12}{l^{4}}+6\frac{\beta^{2}}{r^{6}}.
\end{equation}
in which $\beta=\frac{4q^{2}}{3}.$ It is clear that $r=0$ is a typical
central curvature singularity. According to the values of $\Lambda$\ , $m$
and $q$, this singularity may be clothed by single or double horizons. (See
the reference \cite{MC} for details).

However, for specific values of $\Lambda$\ , $m$ and $q$ the central
curvature singularity becomes \ naked and it deserves to be investigated
within the framework of quantum mechanics. To find the condition for naked
singularities the metric function is written in the following form,

\begin{equation}
f(r)=-\frac{m}{r}\left( r+\tilde{\Lambda}r^{3}-\frac{4\tilde{q}^{2}}{3}%
\right) ,
\end{equation}
where $\tilde{\Lambda}=\frac{\Lambda}{m}$ and $\tilde{q}^{2}=\frac{q^{2}}{m}%
. $ Since the range of coordinate $r$ varies from 0 to infinity, the
negative root will indicate the condition for a naked singularity. In order
to find the roots, we set $f(r)=0$ which yields $r^{3}+\frac{r}{\tilde{%
\Lambda}}-\frac{4\tilde{q}^{2}}{3\tilde{\Lambda}}=0.$ The standard procedure
is followed for a solution via a new variable defined by $r=z-\frac{1}{3%
\tilde{\Lambda}z}$ that transforms the equation to $27\tilde{\Lambda}%
^{3}z^{6}-36\tilde{\Lambda}^{2}\tilde{q}^{2}z^{3}-1=0.$ This equation can be
solved easily and the final answer is

\begin{equation}
r=u^{1/3}-\frac{1}{3\tilde{\Lambda}u^{1/3}},
\end{equation}
in which $u=\frac{12\tilde{q}^{2}\tilde{\Lambda}\pm2\sqrt{3\tilde{\Lambda }%
\left( 12\tilde{q}^{4}\tilde{\Lambda}+1\right) }}{18\tilde{\Lambda}^{2}},$
with a constraint condition $3\tilde{\Lambda}\left( 12\tilde{q}^{4}\tilde{%
\Lambda}+1\right) >0.$ After some algebra, we end up with the following
equation

\begin{equation}
r=a^{1/3}\left\{ \left( 1\pm\frac{b}{a}\right) ^{1/3}+\left( 1\mp\frac {b}{a}%
\right) ^{1/3}\right\} ,
\end{equation}
where $a=\frac{2\tilde{q}^{2}}{3\tilde{\Lambda}}$ and $b=\frac{\sqrt {3%
\tilde{\Lambda}\left( 12\tilde{q}^{4}\tilde{\Lambda}+1\right) }}{9\tilde{%
\Lambda}^{2}}.$ It can be verified easily that the expression inside the
curly bracket in Eq. (18) is always positive. Hence, the only possibility
for a negative root is $a<0$. This implies $\tilde{\Lambda}<0.$ Therefore,
the condition $12\tilde{q}^{4}\tilde{\Lambda}+1<0$ is imposed from the
constraint condition. As a result, for a naked singularity, $\tilde{\Lambda}%
<-\frac {1}{12\tilde{q}^{4}}$ or $\Lambda<-\frac{m^{3}}{12q^{4}}$ should be
satisfied.

Our aim now is to investigate the quantum singularity structure of the naked
singularity that may arise if the constant coefficients satisfy $\Lambda <-%
\frac{m^{3}}{12q^{4}}.$

\subsection{Klein-Gordon Fields:}

Using seperation of variables, $\psi=R(r)e^{in\theta}$, we obtain the radial
portion of Eq. (6) as

\begin{equation}
R_{n}^{\prime\prime}+\frac{\left( fr\right) ^{\prime}}{fr}R_{n}^{\prime }-%
\frac{n^{2}}{fr^{2}}R_{n}-\frac{M^{2}}{f}R_{n}\pm\frac{i}{f^{2}}R_{n}=0,
\end{equation}
where a prime denotes derivative with respect to $r$.

\subsubsection{The case of $r\rightarrow\infty:$}

The Coulomb-like field in metric function (14) becomes negligibly small and
hence, the metric takes the form

\begin{equation}
ds^{2}\simeq-\left( \frac{r^{2}}{l^{2}}\right) dt^{2}+\left( \frac{r^{2}}{%
l^{2}}\right) ^{-1}dr^{2}+r^{2}d\theta^{2}.
\end{equation}

\bigskip

This particular case overlaps with the results already reported in \cite{PL}%
. Hence, no new result arises for this particular case. This is expected
because the effect of source term vanishes for large values of $r$.

\subsubsection{The case of $r\rightarrow0:$}

The case near origin is topologically different compared to the analysis
reported in \cite{PL}. Here, the spacetime is not conic. The approximate
metric near origin is given by,

\begin{equation}
ds^{2}\simeq-\left( \frac{\beta}{r}\right) dt^{2}+\left( \frac{\beta}{r}%
\right) ^{-1}dr^{2}+r^{2}d\theta^{2}.
\end{equation}
This metric can also be interpreted as the $2+1$ dimensional topological
Schwarzchild-like black hole geometry.

For the solution of the radial equation (19), we assume that a massless case
(i.e. $M=0$), and ignore the term $\pm i\frac{R_{n}}{f^{2}}$ ( since it is
negligible near the origin). Then it takes the form,

\begin{equation}
R_{n}^{\prime\prime}-\frac{n^{2}}{\beta r}R_{n}=0,
\end{equation}
whose solution is

\begin{equation}
R_{n}(r)=C_{1n}\sqrt{r}I_{1}(k)+C_{2n}\sqrt{r}K_{1}(k).
\end{equation}
where $I_{1}(k)$ and $K_{1}(k)$ are the first and second kind modified
Bessel functions and $k=\sqrt{\frac{4n^{2}r}{\beta}}.$ The behaviour of the
modified Bessel functions for real $\nu\geq0$ as $r\rightarrow0$ are given
by;

\begin{align}
I_{\nu}(x) & \simeq\frac{1}{\Gamma\left( \nu+1\right) }\left( \frac{x}{2}%
\right) ^{\nu}, \\
K_{\nu}(x) & \simeq\QATOPD{\{}{\}}{-\left[ \ln\left( \frac{x}{2}\right)
+0.5772...\right] ,\text{ \ \ \ \ }\nu=0}{\frac{\Gamma\left( \nu\right) }{2}%
\left( \frac{2}{x}\right) ^{\nu},\text{ \ \ \ \ \ \ \ \ \ \ \ \ \ \ \ \ \ \ }%
\nu\text{\ }\neq0},  \notag
\end{align}
thus $I_{1}(k)\sim\frac{1}{\Gamma\left( 2\right) }\left( \frac{k}{2}\right) $
and $K_{1}(k)\sim\frac{\Gamma\left( 1\right) }{2}\left( \frac{2}{k}\right) $ 
$.$ Checking for the square integrability of the solution (23) requires the
behaviour of the integral for $I_{1}(k)$ $\approx\int r^{4}dr$ and $%
K_{1}(k)\approx\int dr$ which are both convergent as $r\rightarrow0$. Any
linear combination is also square integrable. It follows the solution (23)
belonging to the Hilbert space $\mathcal{H}$ and therefore the operator $A$
described in Eq.(6) is not essentially self-adjoint. So, the naked
singularity at \ $r=0$ is quantum mechanically singular if it is probed with
quantum particles.

Another approach to remove the quantum singularity is to choose the function
space to be the Sobolev space $(H^{1})$ which is used first time in this
context by Ishibashi and Hosoya \cite{IH}. Here, the function space is
defined by $\mathcal{H}=\{R\mid\parallel R\parallel<\infty\},$ where the
norm defined in $2+1$ dimensional geometry as,

\begin{equation}
\parallel R\parallel^{2}\sim\int rf^{-1}\mid R\mid^{2}dr+\int rf\mid \frac{%
\partial R}{\partial r}\mid^{2}dr,
\end{equation}
which involves both the wave function and its derivative to be square
integrable. The failure in the square integrability indicates that the
operator $A$ is essentially self-adjoint and thus, the spacetime is "wave
regular". According to this norm, the first integral is square integrable
while the second integral behaves for the functions $I_{1}(k)$ as $\approx
\int_{0}dr$ and $K_{1}(k)$ integral vanishes. As a result, the wave
functions are square integrable and thus the spacetime is quantum
mechanically wave singular. It should be noted that the Sobolev space is not
the natural quantum mechanical Hilbert space.

\subsection{Dirac Fields:}

We apply the same methodology as in \cite{PL} for finding a solution to
Dirac equation. Since the fermions have only one spin polarization in $2+1$
dimensions \cite{GG}, Dirac matrices are reduced to Pauli matrices \cite{NU}
so that,

\begin{equation}
\gamma^{\left( j\right) }=\left( \sigma^{\left( 3\right) },i\sigma ^{\left(
1\right) },i\sigma^{\left( 2\right) }\right) ,
\end{equation}
where latin indices represent internal (local) indices. In this way,

\begin{equation}
\left\{ \gamma^{\left( i\right) },\gamma^{\left( j\right) }\right\}
=2\eta^{\left( ij\right) }I_{2\times2},
\end{equation}
where $\eta^{\left( ij\right) }$\ is the Minkowski metric in $2+1$\
dimensions and $I_{2\times2}$\ is the identity matrix. The coordinate
dependent metric tensor $g_{\mu\nu}\left( x\right) $\ and matrices $%
\sigma^{\mu}\left( x\right) $\ are related to the triads $e_{\mu}^{\left(
i\right) }\left( x\right) $\ by

\begin{align}
g_{\mu\nu}\left( x\right) & =e_{\mu}^{\left( i\right) }\left( x\right)
e_{\nu}^{\left( j\right) }\left( x\right) \eta_{\left( ij\right) }, \\
\sigma^{\mu}\left( x\right) & =e_{\left( i\right) }^{\mu}\gamma^{\left(
i\right) },  \notag
\end{align}
where $\mu$\ and $\nu$\ are the external (global) indices.

The \ Dirac equation in $2+1$ dimensional curved spacetime for a free
particle with mass $M$\ becomes

\begin{equation}
i\sigma^{\mu}\left( x\right) \left[ \partial_{\mu}-\Gamma_{\mu}\left(
x\right) \right] \Psi\left( x\right) =M\Psi\left( x\right) ,
\end{equation}
where $\Gamma_{\mu}\left( x\right) $\ is the spinorial affine connection and
is given by

\begin{equation}
\Gamma_{\mu}\left( x\right) =\frac{1}{4}g_{\lambda\alpha}\left[ e_{\nu,\mu
}^{\left( i\right) }(x)e_{\left( i\right) }^{\alpha}(x)-\Gamma_{\nu\mu
}^{\alpha}\left( x\right) \right] s^{\lambda\nu}(x),
\end{equation}

\begin{equation}
s^{\lambda\nu}(x)=\frac{1}{2}\left[ \sigma^{\lambda}\left( x\right)
,\sigma^{\nu}\left( x\right) \right] .
\end{equation}

The causal structure of the spacetime indicates that there are two singular
cases to be investigated. The asymptotic case, $r\rightarrow\infty$ has
already been analysed by PL. The case of $r\rightarrow0$ is not conical so
there is a topological difference in the spacetime near $r=0$ . Hence, the
suitable triads for the metric (21) are given by,

\begin{equation}
e_{\mu}^{\left( i\right) }\left( t,r,\theta\right) =diag\left( \left( \frac{%
\beta}{r}\right) ^{\frac{1}{2}},\left( \frac{r}{\beta}\right) ^{\frac{1}{2}%
},r\right) ,
\end{equation}
The coordinate dependent gamma matrices and the spinorial affine connection
are given by

\begin{align}
\sigma^{\mu}\left( x\right) & =\left( \left( \frac{r}{\beta}\right) ^{\frac{1%
}{2}}\sigma^{\left( 3\right) },i\left( \frac{\beta}{r}\right) ^{\frac{1}{2}%
}\sigma^{\left( 1\right) },\frac{i\sigma^{\left( 2\right) }}{r}\right) , \\
\Gamma_{\mu}\left( x\right) & =\left( \frac{-\beta\sigma^{\left( 2\right) }}{%
4r^{2}},0,\frac{i}{2}\left( \frac{\beta}{r}\right) ^{\frac {1}{2}%
}\sigma^{\left( 3\right) }\right) .  \notag
\end{align}
Now, for the spinor

\begin{equation}
\Psi=\left( 
\begin{array}{c}
\psi_{1} \\ 
\psi_{2}%
\end{array}
\right) ,
\end{equation}
the Dirac equation can be written as

\begin{align}
i\left( \frac{r}{\beta}\right) ^{\frac{1}{2}}\frac{\partial\psi_{1}}{%
\partial t}-\left( \frac{\beta}{r}\right) ^{\frac{1}{2}}\frac{\partial
\psi_{2}}{\partial r}+\frac{i}{r}\frac{\partial\psi_{2}}{\partial\theta}-%
\frac{1}{4}\left( \frac{\beta}{r^{3}}\right) ^{\frac{1}{2}%
}\psi_{2}-M\psi_{1} & =0, \\
-i\left( \frac{r}{\beta}\right) ^{\frac{1}{2}}\frac{\partial\psi_{2}}{%
\partial t}-\left( \frac{\beta}{r}\right) ^{\frac{1}{2}}\frac{\partial
\psi_{1}}{\partial r}-\frac{i}{r}\frac{\partial\psi_{1}}{\partial\theta}-%
\frac{1}{4}\left( \frac{\beta}{r^{3}}\right) ^{\frac{1}{2}%
}\psi_{1}-M\psi_{2} & =0.  \notag
\end{align}
The following ansatz will be employed for the positive frequency solutions:

\begin{equation}
\Psi_{n,E}\left( t,x\right) =\left( 
\begin{array}{c}
R_{1n}(r) \\ 
R_{2n}(r)e^{i\theta}%
\end{array}
\right) e^{in\theta}e^{-iEt}.
\end{equation}
The radial parts of the Dirac equation for investigating the behaviour as $%
r\rightarrow0$, are

\begin{align}
R_{1n}^{\prime\prime}+\frac{\alpha_{1}}{\sqrt{r}}R_{1n}^{\prime}+\frac {%
\alpha_{2}}{r^{3/2}}R_{1n} & =0, \\
R_{2n}^{\prime\prime}+\frac{\alpha_{3}}{\sqrt{r}}R_{2n}^{\prime}+\frac {%
\alpha_{4}}{r^{3/2}}R_{2n} & =0.  \notag
\end{align}
where $\alpha_{1}=\frac{2M-E}{2M\sqrt{\beta}},$ $\alpha_{2}=\frac {%
-7E+4M\left( 4n+1\right) }{16M\sqrt{\beta}}$, $\alpha_{3}=\frac {2M+E}{2M%
\sqrt{\beta}}$ and $\alpha_{4}=\frac{7E-4M\left( 4n+3\right) }{16M\sqrt{\beta%
}}.$ Then, the solutions are given by;

\begin{align*}
R_{1}(r) & =e^{-\frac{b}{2}\rho}\left\{ C_{1}\sqrt{\rho}\text{\textrm{%
Whittaker}}_{M}(a,1,b\rho)+C_{2}\sqrt{\rho}\text{\textrm{Whittaker}}%
_{W}(a,1,b\rho)\right\} , \\
R_{2}(r) & =e^{-\frac{b^{\prime}}{2}\rho}\left\{ C_{3}\sqrt{\rho }\text{%
\textrm{Whittaker}}_{M}(a^{\prime},1,b^{\prime}\rho)+C_{4}\sqrt{\rho }\text{%
\textrm{Whittaker}}_{W}(a^{\prime},1,b^{\prime}\rho)\right\}
\end{align*}
where $\rho=\sqrt{r}$, $a=\frac{-9E+8M(1+2n)}{4(2M-E)},a^{\prime}=\frac{%
9E-8M(1+2n)}{4(2M+E)},b=2\alpha_{1},$and $b^{\prime}=2\alpha_{3}$.

When we look for the square integrability of the above solutions, we
obtained that both functions \textrm{Whittaker}$_{M}$ and \textrm{Whittaker}$%
_{W}$ are square integrable near $\rho=0$ (or $r=0$) for both $R_{1}(r)$ and 
$R_{2}(r)$. One has,%
\begin{equation}
\int rf^{-1}\mid R\mid^{2}dr\approx\int\rho^{6}e^{-b\rho}\left[ \mathrm{%
Whittaker}_{M}(a,1,b\rho)\right] ^{2}d\rho<\infty,
\end{equation}

\bigskip and

\begin{equation}
\approx\int\rho^{6}e^{-b\rho}\left[ \mathrm{Whittaker}_{W}(a,1,b\rho)\right]
^{2}d\rho<\infty.
\end{equation}
We note that these results are verified first by expanding the Whittaker
functions in series form up to the order of $\mathcal{O(\rho}^{6})$ and then
by integrating term by term in the limit as $r\rightarrow0$.

$\bigskip$The set of solutions for the Dirac equation for the spacetime (21)
is given by

\begin{equation*}
\Psi_{n,E}\left( t,\mathbf{x}\right) =\left( 
\begin{array}{c}
e^{-\frac{b}{2}\rho}\left\{ C_{1n}\sqrt{\rho}\mathrm{Whittaker}%
_{M}(a,1,b\rho)+C_{2n}\sqrt{\rho}\mathrm{Whittaker}_{W}(a,1,b\rho)\right\}
\\ 
e^{-\frac{b^{\prime}}{2}\rho}\left\{ C_{3n}\sqrt{\rho}\mathrm{Whittaker}%
_{M}(a^{\prime},1,b^{\prime}\rho)+C_{4n}\sqrt{\rho}\mathrm{Whittaker}%
_{W}(a^{\prime},1,b^{\prime}\rho)\right\} e^{i\theta}%
\end{array}
\right) e^{in\theta}e^{-iEt},
\end{equation*}
and an arbitrary wave packet can be written as

\begin{equation}
\Psi \left( t,\mathbf{x}\right) =\overset{+\infty }{\underset{n=-\infty }{%
\dsum }}C_{n}\left( 
\begin{array}{c}
e^{-\frac{b}{2}\rho }\sqrt{\rho }(\mathrm{Whittaker}_{M}(a,1,b\rho )+\mathrm{%
Whittaker}_{W}(a,1,b\rho )) \\ 
e^{-\frac{b^{\prime }}{2}\rho }\sqrt{\rho }(\mathrm{Whittaker}_{M}(a^{\prime
},1,b^{\prime }\rho )+\mathrm{Whittaker}_{W}(a^{\prime },1,b^{\prime }\rho
))e^{i\theta }%
\end{array}%
\right) e^{in\theta }e^{-iEt}
\end{equation}%
where $C_{n}$ is an arbitrary constant. Hence, initial condition $\Psi
\left( 0,\mathbf{x}\right) $ is sufficient to determine the future time
evolution of the particle. The spacetime is then quantum regular when tested
by fermions.

\section{$\left( 2+1\right) $ - Dimensional BTZ Spacetime With Linear
Electrodynamics}

\subsection{Solutions and Spacetime Structure:}

The metric for the charged BTZ spacetime in linear electrodynamics is given
by \cite{MZ},

\begin{equation}
ds^{2}=-f(r)dt^{2}+f(r)^{-1}dr^{2}+r^{2}d\theta^{2},
\end{equation}
with the metric function

\begin{equation}
f(r)=-m+\frac{r^{2}}{l^{2}}-2q^{2}\ln(\frac{r}{l}),
\end{equation}
where $q$ is the electric charge and $m>0$ is the mass and $l^{2}=\Lambda
^{-1}$\ . The Kretschmann scalar is given by,

\begin{equation}
\mathcal{K}=\frac{12}{l^{4}}-\frac{8q^{2}}{r^{2}l^{2}}+\frac{4q^{4}}{r^{4}},
\end{equation}
which displays a power-law central curvature singularity at $r=0$. According
to the values of $m$, $l$ and $q,$ this central singularity is clothed by
horizons or it remains naked. Our interest here is to investigate the
quantum mechanical behaviour of the naked singularity. In order to find the
condition for naked singularity, we set $f(r_{h})=0$ and the solution for $%
l=1$ is

\begin{equation*}
r_{h}=\exp \left\{ -\frac{m}{2q^{2}}-\frac{1}{2}\text{\textrm{LambertW}}%
\left( -\frac{1}{q^{2}}e^{-m/q^{2}}\right) \right\} ,
\end{equation*}%
in which \textrm{LambertW(.) }represents the Lambert function \cite{LM}.%
\textrm{\ }Fig.1 displays (unmarked region) the possible values of $m$ and $%
q $ that result in naked singularity.

The causal structure is similar to the case considered in the previous
section. There are two singular cases to be investigated. The case for $%
r\rightarrow \infty $ is approximately the same case considered in \cite{PL}%
. Therefore, the results reported by PL are valid for this case as well. For
small $r$ values, the approximate metric can be written in the following form

\begin{equation}
ds^{2}\approx -\left( 2q^{2}\mid \ln \left( \tilde{r}\right) \mid \right)
dt^{2}+\left( 2q^{2}\mid \ln \left( \tilde{r}\right) \mid \right)
^{-1}dr^{2}+r^{2}d\theta ^{2},
\end{equation}%
in which $\tilde{r}=\frac{r}{l}<<1.$

\subsection{Klein-Gordon Fields:}

The radial equation for the metric (44) is obtained for the massless case as,

\begin{equation}
R_{n}^{\prime\prime}+\frac{1}{\tilde{r}}\left( 1+\frac{1}{\ln\tilde{r}}%
\right) R_{n}^{\prime}+\frac{n^{2}}{2q^{2}r^{2}\ln\tilde{r}}R_{n}=0,
\end{equation}
since $\frac{r}{l}<<1$, the solution can be written in terms of zeroth order
first and second kind modified Bessel functions,

\begin{equation}
R_{n}(x)=C_{1n}I_{0}(\frac{\sqrt{2}n}{q}x)+C_{2n}K_{0}(\frac{\sqrt{2}n}{q}x),
\end{equation}
where $-x^{2}=\ln\tilde{r}$. As $\tilde{r}\rightarrow0$, $%
x\rightarrow\infty. $ The behaviour of the modified Bessel functions for $%
x>>1$ are $I_{0}\left( x\right) \simeq\frac{e^{x}}{\sqrt{2\pi x}}$ and $%
K_{0}\left( x\right) \simeq\sqrt{\frac{\pi}{2x}}e^{-x}.$ These functions are
always square integrable for $x\rightarrow\infty,$ that is

\begin{equation*}
\int rf^{-1}\mid R\mid^{2}dr\approx\int xe^{-2x^{2}}f^{-1}\mid R\mid
^{2}dx<\infty.
\end{equation*}

These results indicate that charged BTZ black hole in linear electrodynamics
is quantum mechanically singular when probed with quantum test particles
that obey Klein-Gordon equation.

If we use the Sobolev norm (25), the second integral which involves the
derivative of the wave function $I_{0}\left( x\right) \simeq\frac{e^{x}}{%
\sqrt{2\pi x}}$ becomes $\approx\int x^{-2}e^{2x}\left( 2x-1\right) ^{2}dx.$
Numerical integration has revealed that as $x\rightarrow\infty,$ $\sim\int
x^{-2}e^{2x}\left( 2x-1\right) ^{2}dx\rightarrow\infty.$ On the other hand
for the wave function $K_{0}\left( x\right) \simeq\sqrt{\frac {\pi}{2x}}%
e^{-x}$, the second integral in the Sobolev norm is solved numerically as $%
x\rightarrow\infty,$ $\sim\int x^{-2}e^{-2x}\left( 2x+1\right) ^{2}dx$ $%
<\infty$ which is square integrable. As a result, charged coupled BTZ black
hole in linear electrodynamics is quantum mechanically wave regular if and
only if the arbitrary constant parameter is $C_{2n}=0$ in Eq.(46).

Consequently, if the naked singularity both in linear and non-linear
electrodynamics is probed with quantum test particles, the following results
are obtained:

1) In classical point of view, the Kretschmann scalar in non-linear case
diverges faster than in the linear case.

2) In quantum mechanical point of view, if the chosen function space is
Sobolev space, spacetime remains singular for non-linear case, but the
spacetime can be made wave regular for linear case.

From these results we may conclude that the structure of the naked
singularity in the non-linear electrodynamics is deeper rooted than the
singularity in the linear case.

\subsection{Dirac Fields:}

The effect of the charge when $r\rightarrow\infty$ does not contribute as
much as the term that contains the cosmological constant. Therefore, we
ignore the mass and the charged terms in the metric function (42). This
particular case has already been analysed in \cite{PL}. The contribution of
the charge is dominant when $r\rightarrow0.$ The Dirac equation for the
metric (44) is solved by using the same method demonstrated in the previous
section. We obtain the radial equation in the limit $r\rightarrow0$ as

\begin{equation}
R_{j}^{\prime\prime}+\frac{1}{r}R_{j}^{\prime}-\frac{R_{j}}{4r^{2}}=0,\text{
\ \ \ \ \ \ \ \ \ \ \ \ \ \ \ }j=1,2
\end{equation}
whose solution is given by

\begin{equation}
R_{j}(r)=C_{1j}\sqrt{r}+\frac{C_{2j}}{\sqrt{r}}.
\end{equation}
where $C_{1j}$ and $C_{2j}$ are arbitrary constants. The solution given in
Eq.(48) is square integrable. The arbitrary wave packet can be written as,

\begin{equation}
\Psi\left( t,\mathbf{x}\right) =\overset{+\infty}{\underset{n=-\infty}{\dsum 
}}\left( \QATOPD{\{}{\}}{R_{1}\left( r\right) }{R_{2}\left( r\right)
e^{i\theta}}\right) e^{in\theta}e^{-iEt}
\end{equation}
Thus, the spacetime is quantum mechanically regular when probed with
fermions.

\section{$\left( 2+1\right) $ - Dimensional Einstein - Maxwell - Dilaton
Gravity}

\subsection{Solutions and Spacetime Structure:}

In this section, we consider $3D$ black holes described by the
Einstein-Maxwell-dilaton action,

\begin{equation}
S=\int d^{3}x\sqrt{-g}\left( R-\frac{B}{2}\left( \bigtriangledown
\phi\right) ^{2}-e^{-4a\phi}F_{\mu\nu}F^{\mu\nu}+2e^{b\phi}\Lambda\right) ,
\end{equation}
where $R$ is the Ricci scalar, $\phi$ is the dilaton field, $F_{\mu\nu}$ is
the Maxwell field and $\Lambda$, $a$, $b$, and $B$ are arbitrary couplings.
The general solution to this action is given by \cite{CM},

\begin{equation}
ds^{2}=-f(r)dt^{2}+\frac{4r^{\frac{4}{N}-2}dr^{2}}{N^{2}\gamma^{\frac{4}{N}%
}f(r)}+r^{2}d\theta^{2},
\end{equation}
where

\begin{equation}
f(r)=Ar^{\frac{2}{N}-1}+\frac{8\Lambda r^{2}}{\left( 3N-2\right) N}+\frac{%
8Q^{2}}{(2-N)N}.
\end{equation}
Here, $A$ is a constant of integration which is proportional to the
quasilocal mass $(A=\frac{-2m}{N})$, $\gamma$\ is an integration constant \
and $Q$ is the charge. The dilaton field is given by

\begin{equation}
\phi=\frac{2k}{N}\ln\left( \frac{r}{\beta\left( \gamma\right) }\right)
\end{equation}
in which $\beta\left( \gamma\right) $ is a $\gamma$ related constant
parameter. Note that, the above solution for $N=2$ contains both the vacuum
BTZ metric if one takes $Q=A=0$ and the BTZ black hole if $A<0,Q=0$. \
However, if the constant parameters are chosen appropriately, the resulting
metric represents black hole solutions with prescribed properties. For
example, when $N=\frac{6}{5}$, $A=-\frac{5m}{3},$ the metric function given
in equation (52) becomes

\begin{equation}
f(r)=-\frac{5m}{3}r^{2/3}+\frac{25\Lambda}{6}r^{2}+\frac{25Q^{2}}{3},
\end{equation}
and therefore the corresponding metric is

\begin{equation}
ds^{2}=-f(r)dt^{2}+\frac{\alpha r^{4/3}dr^{2}}{f(r)}+r^{2}d\theta^{2},
\end{equation}
where $\alpha=\frac{25}{9\gamma^{\frac{10}{3}}}$ is a constant parameter.

\bigskip The Kretschmann scalar for this solution is given by

\begin{equation}
\mathcal{K}=\frac{25\left\{ 12m^{2}r^{\frac{5}{3}}+5\Lambda r^{3}\left[
55\Lambda r^{\frac{4}{3}}-4m\right] +40r^{\frac{1}{3}}Q^{2}\left[ 2\left(
5Q^{2}-mr^{\frac{2}{3}}\right) -5\Lambda r^{2}\right] \right\} }{%
81\alpha^{2}r^{7}},
\end{equation}
which indicates a central curvature singularity at $r=0$ that is clothed by
the event horizon. To find the location of horizons, $g_{tt}$ is set to zero
and we have

\bigskip

\begin{equation}
r^{2}-\frac{2m}{5\Lambda}r^{\frac{2}{3}}+\frac{2Q^{2}}{\Lambda}=0.
\end{equation}

There are three possible cases to be considered.

Case 1: If $\frac{Q^{2}}{\Lambda}<\left( \frac{2m}{15\Lambda}\right) ^{\frac{%
3}{2}}$, the equation admits two positive roots indicating inner and outer
horizons of the black hole.

Case 2: If $\frac{Q^{2}}{\Lambda}=\left( \frac{2m}{15\Lambda}\right) ^{\frac{%
3}{2}}$, this is an extreme case and the equation (57) has one real positive
root. This means that there is only one horizon.

Case 3: If $\frac{Q^{2}}{\Lambda}>\left( \frac{2m}{15\Lambda}\right) ^{\frac{%
3}{2}},$\ there is no real positive root and the solution does not admit
black hole so that the singularity at $r=0$ is naked. With reference to the
detailed analysis given in \cite{CM}, the Penrose diagram of the solution
illustrates the timelike character of the singularity at $r=0$. Our aim in
this section is to investigate the behaviour of this naked singularity when
probed with Klein-Gordon and Dirac fields in the framework of quantum
mechanics.

\subsection{Klein-Gordon Fields:}

The radial equation for the metric (55) is obtained for the massles case$%
(M=0)$ as,

\begin{equation}
R_{n}^{\prime\prime}+\frac{\left( fr^{\frac{1}{3}}\right) ^{\prime}}{fr^{%
\frac{1}{3}}}R_{n}^{\prime}-\frac{\alpha n^{2}}{fr^{\frac{2}{3}}}R_{n}\pm%
\frac{i\alpha r^{\frac{4}{3}}}{f^{2}}R_{n}=0.
\end{equation}
The behaviour of the radial equation as $r\rightarrow0$ is

\bigskip

\begin{equation}
R_{n}^{\prime\prime}+\frac{1}{3r}R_{n}^{\prime}-\frac{k^{2}}{r^{\frac{2}{3}}}%
R_{n}=0,
\end{equation}
where $k=\frac{3\alpha n^{2}}{25Q^{2}}.$ \ The solution is given by

\begin{equation}
R_{n}(r)=C_{1n}\cosh\left( \frac{3k}{2}r^{2/3}\right) +iC_{2n}\sinh\left( 
\frac{3k}{2}r^{2/3}\right) .
\end{equation}
Both solutions are square integrable in Hilbert space, that is, $\int
rg_{rr}\mid R\mid^{2}dr<\infty.$ Therefore, the spacetime is quantum
mechanically singular when probed with quantum particles obeying
Klein-Gordon equation.

If we use the Sobolev norm,

\begin{equation*}
\parallel R\parallel^{2}\sim\int rg_{rr}\mid R\mid^{2}dr+\int
rg_{rr}^{-1}\mid\frac{\partial R}{\partial r}\mid^{2}dr,
\end{equation*}
although the first integral of the solution is square integrable, the second
integral for $C_{1n}=0$ fails to be square integrable and the spacetime is
quantum mechanically wave regular.

\subsection{Dirac Fields:}

The Dirac equation can be written as,

\bigskip%
\begin{align}
\frac{i}{\sqrt{f}}\psi_{1,t}-\frac{\sqrt{f}}{\sqrt{\alpha}r^{\frac{2}{3}}}%
\psi_{2,r}+\frac{i}{r}\psi_{2,\theta}-\left\{ \frac{5(-2m+15\Lambda r^{\frac{%
4}{3}})}{36\sqrt{\alpha}r\sqrt{f}}+\frac{\sqrt{f}}{2\sqrt{\alpha }r^{\frac{5%
}{3}}}\right\} \psi_{2}-M\psi_{1} & =0, \\
\frac{-i}{\sqrt{f}}\psi_{2,t}-\frac{\sqrt{f}}{\sqrt{\alpha}r^{\frac{2}{3}}}%
\psi_{1,r}-\frac{i}{r}\psi_{1,\theta}-\left\{ \frac{5(-2m+15\Lambda r^{\frac{%
4}{3}})}{36\sqrt{\alpha}r\sqrt{f}}+\frac{\sqrt{f}}{2\sqrt{\alpha }r^{\frac{5%
}{3}}}\right\} \psi_{1}-M\psi_{2} & =0  \notag
\end{align}
where $f$ is given in (54). By using the same anzats as in (36), \ the
radial part of the Dirac equation becomes,

\begin{equation}
R_{n}^{\prime\prime}+\frac{a_{1}}{r^{1/3}}R_{n}^{\prime}+\frac{a_{2}}{r^{2/3}%
}R_{n}=0,\text{ \ \ \ \ \ \ \ \ \ }n=1,2
\end{equation}
in which $a_{1}=\frac{3Q\sqrt{3\alpha}-m}{15Q^{2}}$, $a_{2}=\frac {%
-108Q^{2}\alpha n\left( 1+n\right) +m\left( m-6Q\sqrt{3\alpha}\right) }{%
900Q^{4}}.$ The solution becomes,

\begin{equation}
R_{n}(r)=r^{1/6}e^{-\frac{3a_{1}}{4}r^{2/3}}\left\{ 
\begin{array}{c}
C_{1n}\mathrm{Whittaker}_{M}(\frac{a_{1}}{4\sqrt{a_{1}^{2}-4a_{2}}},\frac {3%
}{4},\frac{3}{2}r^{2/3}\sqrt{a_{1}^{2}-4a_{2}})+ \\ 
C_{2n}\mathrm{Whittaker}_{W}(\frac{a_{1}}{4\sqrt{a_{1}^{2}-4a_{2}}},\frac {3%
}{4},\frac{3}{2}r^{2/3}\sqrt{a_{1}^{2}-4a_{2}})%
\end{array}
\right\} ,
\end{equation}
which is square integrable. This is verified first by expanding the
Whittaker functions in series and then by integrating term by term in the
limit as $r\rightarrow0.$ Consequently, the spacetime is quantum
mechanically regular when probed with Dirac fields.

\section{Conclusion}

Matter coupled $2+1$ dimensional black hole spacetimes are shown to share
similar quantum mechanical singularity structure as in the case of pure BTZ
black hole. The inclusion of matter fields changes the topology and creates
true curvature singularity at $r=0$. The effect of the matter fields allows
only specific frequency modes in the solution of Klein-Gordon and Dirac
fields. If the quantum singularity analysis is based on the natural Hilbert
space of quantum mechanics which is the linear function space with square
integrability $L^{2}$, the singularity at $r=0$ turns out to be quantum
mechanically singular for particles obeying the Klein- Gordon equation and
regular for fermions obeying the Dirac equation. We have proved that the
quantum singularity structure of $2+1$ dimensional black hole spacetimes are
generic for Dirac particles and the character of the singularity in quantum
mechanical point of view is irrespective whether the matter field is coupled
or not. This result suggests that the Dirac fields preserve the cosmic
censorhip hypothesis in the considered spacetimes that exhibit timelike
naked singularities. Instead of horizons (that clothes the singularity in
the black hole cases) the repulsive barrier is replaced against the
propagation of Dirac fields. However, for particles obeying Klein - Gordon
fields, the singularity becomes worse when a matter field is coupled.

However, we have also shown that in the charged BTZ ( in linear
electrodynamics ) and dilaton coupled black hole spacetimes specific choice
of waves exhibit quantum mechanical wave regularity when probed with waves
obeying Klein-Gordon equation, if the function space is Sobolev with the
norm defined in (25). The singularity at $r=0$ is stronger in the non-linear
electrodynamic case. It should be reminded that, one may not feel
comfortable to use Sobolev norm in place of natural linear function space of
quantum mechanics.

\

\begin{center}
\textbf{Figure Captions}
\end{center}

Figure1: Plot of $r_{h}$ for different values of $m$ and $q$. Marked region
displays the formation of black hole, unmarked region shows the formation of
naked singularity.

\end{document}